\documentstyle[aps,amsfonts,twocolumn,prbbib]{revtex}

\begin{document}
 
 \title{Pairs of Bloch electrons and magnetic translation groups}
 \author{Wojciech Florek\cite{AAA}}
 \address{A.~Mickiewicz University, Institute of Physics, ul.~Umultowska
85, 61-614 Pozna\'n, Poland}
 \date{\today}

\maketitle

\begin{abstract}
 A product of irreducible representations of magnetic translation group is
considered. It leads to irreducible representations which were previously
rejected as `nonphysical' ones. A very simple example indicates a possible
application of these representations. In particular, they are important in
descriptions of pairs of electrons in a magnetic field and a periodic
potential. The periodicity of some properties with respect to the charge of
a particle is briefly discussed.
 \end{abstract}


\section{Introduction}
 The first attempts to describe movement of electrons in the presence of a
constant external magnetic field had been done by Landau \cite{landau} and
Peierls. \cite{peierls} In the fifties many authors dealt with similar
problems, but a crystalline (periodic) potential was also included.
\cite{many} The pioneering works of Brown \cite{brown} and Zak
\cite{zak1,zak2} were preceded by Wannier's paper. \cite{wannier} The first
two authors independently introduced and investigated the so-called {\em
magnetic translations}, i.e.\ unitary, mutually {\em noncommuting},
operators which commute with the Hamiltonian. For more than thirty years
these operators have been applied in many problems concerning movement of
electrons in crystal lattice.  Recently, much attention has been paid to
two-dimensional systems in the external magnetic field due their relations
with high-$T_c$ superconductors, anyons, the Hall effect etc. \cite{aoki}  

From the group-theoretical point of view magnetic translations can be
considered as a projective (ray) representation of the translation group $T$
of a~crystal lattice (this is Brown's approach). However, projective
representations of any group can be found as vector representations of its
covering group (the so-called {\em magnetic translation group} {\sf MTG}).
This later group can be constructed as a~central extension of a~given group
by the group of factors, in general $U(1)\subset{\Bbb C}^*$ or its subgroup.
This construction is a~basis of Zak's considerations and is very close
related to the Mac~Lane method for determination of all inequivalent
(abelian) extensions of two groups. \cite{torun} In this paper irreducible
representations (irreps) of {\sf MTG}'s are considered. In fact they were
determined by Brown and Zak \cite{brown,zak1} but both authors rejected most
of them as `nonphysical'. It is here shown that all representations are
`physical' and a very simple example of their applications is presented.
Moreover, the Clebsch--Gordan coefficients are calculated in this case.

\section{Magnetic translations}
 The {\sf MTG} appears in a natural way when one considers an
electron in a periodic potential $V({\bf r})$ and a uniform magnetic
field {\bf H} determined by a vector potential (a gauge) {\bf A}.
This system is described by the well-known Hamiltonian
 \begin{equation}\label{hamil}
  {\cal H}\;=\;\frac{1}{2m}({\bf p}+e{\bf A}/c)^2
  +V({\bf r})\,,
 \end{equation}
 which does {\em not}\/ commute with the usual translation operators 
 \begin{equation}\label{ordin}
  \widehat{T}_0({\bf R})\;=\;
  \exp(-{\rm i}{\bf R}\cdot{\bf p}/\hbar)
 \end{equation}
 and 
 \begin{equation}
  \widehat{T}_0({\bf R})\psi({\bf r})\;=\;\psi({\bf r}-{\bf R})
 \end{equation}
 (the subscript `0' corresponds to ${\bf H}=0$). However, the
Hamiltonian (\ref{hamil}) commutes with unitary operators
 \begin{equation}\label{ray}
 \widehat{T}({\bf R})\;=\;\exp[-{\rm i}
 ({\bf p}-e{\bf A}/c)\cdot{\bf R}/\hbar]
 \end{equation}
 introduced by Brown. \cite{brown} It is easy to check that in this
way a projective representation of the translation group is defined;
the corresponding factor system is given as
 \begin{equation}\label{facm}
  m({\bf R},{\bf R}')\;=\;\exp\Bigl[-\frac{1}{2}
  \frac{{\rm i}e}{c\hbar} 
  ({\bf R}\times{\bf R}')\cdot{\bf H}\Bigr].
 \end{equation}
 It has to be stressed that these operators commute with the
Hamiltonian (\ref{hamil}) if the vector potential~{\bf A} fulfills
the condition
 \begin{equation} \label{Ac}
  \partial A_j/\partial x_k
  +\partial A_k/\partial x_j\:=\:0\qquad 
  {\rm for}\quad  j,k\,=\,1,2,3\,.
 \end{equation}
 This relation holds, for example, for the (global) gauge 
 ${\bf A}({\bf r})= \case{1}{2}({\bf H}\times{\bf r})$, which was
used by both authors. \cite{brown,zak1} It is worth noting that
introducing a {\em local}\/ gauge one can consider any vector potential
{\bf A}.
\cite{poz0}

{\em Projective}\/ representations of a given group are related with
{\em vector}\/ representations of the covering group, which can be
determined as a central extension. In the considered case, one deals
with representations of the translation group $T\simeq{\Bbb Z}^3$ and
the magnetic translation group ${\cal T}$ is its covering group,
i.e.\ ${\cal T}$ is included in a central extension of $T$ by $U(1)$.
Let ${\cal T}$ consist of pairs $(u,{\bf R})$, $u\in U(1)$, ${\bf
R}\in T$, with the multiplication rule
 \begin{equation}\label{mulrul}
 (u,{\bf R})(u',{\bf R}')
 \;=\;(uu'\,m({\bf R},{\bf R}'),{\bf R}+{\bf R}')
 \end{equation}
 with $m\colon T\times T\to U(1)$ being a factor system, and let
$\Xi$ be an irrep of $U(1)$. An irrep of ${\cal T}$ is given as a
product
 \begin{equation}\label{irrep}
 \Gamma(u,{\bf R})\;=\;\Xi(u)\Lambda({\bf R})\,,
 \end{equation}
 where  $\Lambda$ is a~projective representation of~$T$ with a~factor
system 
 \begin{equation}\label{nuf} 
  \nu({\bf R},{\bf R}')\;=\;\Xi(m({\bf R},{\bf R}'))\,.
 \end{equation}
 Zak \cite{zak1,zak2} introduced such a covering group by attaching to each
vector {\bf R} a path ${\cal P}$ drawn in the crystal lattice, i.e.\
consisting of vectors ${\bf R}_1,{\bf R}_2,\ldots,{\bf R}_k\in T$, such that
$\sum_{j=1}^k{\bf R}_j={\bf R}$.  The magnetic flux $\Phi$ through the
polygon enclosed by a loop ${\cal L}={\cal P}\cup\{-{\bf R}\}$ determines a
factor $u({\cal L})=\exp(-{\rm i} e\Phi/c\hbar)\in U(1)$.  Pairs $(u({\cal
L}),\widehat{T}_0({\bf R}))$ form the magnetic translation group in Zak's
approach [$\widehat{T}_0$ is defined in Eq.\ (\ref{ordin})].

  Zak showed that a factor system of the covering group he introduced is
identical with the factor system (\ref{facm}). Therefore, both these
approaches are equivalent if $\Xi(u)=u$. Other representations of ${\cal T}$
were rejected by Zak, since they were viewed as `nonphysical'.

{\sf MTG}'s were considered as central extensions in some previous papers
\cite{torun,poz1,poz2} as an illustrative example of the Mac~Lane method for
determination of all inequivalent extensions of given groups (the Mac~Lane
method is discussed in a review article by Lulek \cite{mll}). This algebraic
description allows deep investigations of {\sf MTG}'s and their
representations.

\section{Irreducible representations}
  Let $\Gamma^1$ and $\Gamma^2$ be irreps of ${\cal T}$ satisfying
the condition $\Xi(u)=u$. Matrix elements of their Kronecker product
$\Gamma=\Gamma^1\otimes\Gamma^2$ can be found as
 \begin{equation}\label{prod}
 \Gamma_{jk,lm}(u,{\bf R})\;=\;\Gamma^1_{j,l}
 (u,{\bf R})\Gamma^2_{k,m}(u,{\bf R})\,.
 \end{equation}
 Taking into account the definition of an irrep (\ref{irrep}) one
obtains
 \begin{equation}\label{prodll}
 \Gamma_{jk,lm}(u,{\bf R})\;=\;
 u^2\Lambda^1_{j,l}({\bf R})\Lambda^2_{k,m}({\bf R})\,.
 \end{equation}
 The last product in this formula determines a product
$\Lambda=\Lambda^1\otimes\Lambda^2$ of two projective
representations, which is a projective representation itself. To
determine its factor system one has to calculate a product
$\Lambda({\bf R}_1)\Lambda({\bf R}_2)$:
 \begin{eqnarray*}
 \left(\Lambda({\bf R}_1)\Lambda({\bf R}_2)\right)_{jk,lm}&=&
  \sum_{n,p}\Lambda_{jk,np}({\bf R}_1)\Lambda_{np,lm}({\bf R}_2)\\
  &=& m({\bf R}_1,{\bf R}_2)^2
  \Lambda_{jk,lm}({\bf R}_1+{\bf R}_2)\,.
 \end{eqnarray*}
 Therefore, the representation $\Lambda$ has the factor system $\nu({\bf
R},{\bf R}')\;=m({\bf R},{\bf R}')^2$, which means that it corresponds to
the irrep $\Xi(u)=u^2$ [cf.\ Eq.\ (\ref{prodll})]. In the other words --- a
product of two `physical' representations gives a `nonphysical' one.
However, there are no {\em a priori}\/ rules to exclude (as `nonphysical') a
product of two (`physical') representations. Therefore, it {\em has}\/ to be
assumed that also $\Gamma$ is relevant for physics.  

Zak had rejected irreps with $\Xi(u)\neq u$ since `{\em representations with
the correspondence $\epsilon\to\epsilon^n$ with $n\neq1$ are nonphysical.}'
\cite{zak,zak1a} However, the above mentioned constant contains the electric
charge (of an electron). If one assumes that representations with
$\Xi(u)=u^2$ describe movement of a particle (or a system of particles) with
a charge $Q=-2e$ then all formulae will be consistent. The simplest
interpretation says that such representations describe a {\em pair}\/ of
electrons. This agrees with the way in which they have been obtained: 
$\Gamma$ describing a pair of electrons is a product of two one-electron
representations $\Gamma^1$ and $\Gamma^2$. Writing the Hamiltonian
(\ref{hamil}) in the form
 \begin{equation}\label{hamilp}
  {\cal H}\;=\;\frac{1}{2\beta m}({\bf p}+\alpha e
  {\bf A}/c)^2+V({\bf r})
 \end{equation}
 one can say that for $\alpha=\beta$ it describes movement of
$\alpha$ electrons in the magnetic field and the periodic potential.
If $\beta\neq\alpha=0$ then this Hamiltonian corresponds to a
particle of a mass $\beta m$ without electric charge. Since
$\alpha=0$ then both factor systems (for the central extension ${\cal
T}$ and the projective representation $\widehat{T}$) are trivial and
the original translation group $T$ and its vector irreps are
appropriate to describe dynamics of the system. (The magnetic field
is irrelevant if one considers classical or spinless particles, of
course.)

\section{Finite two-dimensional {\sf MTG}'s}
 One can introduce finite representations of {\sf MTG}'s imposing the
periodic boundary conditions in the form $\widehat{T}(N{\bf a}_j)=1$,
where ${\bf a}_j$, $j=1,2,3$, are the unit vectors of a crystal
lattice. \cite{brown} This is equivalent to considerations of a
finite translation group $T={\Bbb Z}_N^3$ (identical periods in each
direction are assumed). Both approaches yield that the magnetic field
should be parallel to a lattice vector. It is convenient to assume
that ${\bf H}\parallel{\bf a}_3$ and is perpendicular to ${\bf a}_1$
and ${\bf a}_2$.  This allows to consider $T={\Bbb Z}_N^2$ and a
factor group to be $C_N$ (the multiplicative group of the $N$-th
roots of 1). Therefore, a finite two-dimensional magnetic translation
group is a central extension of a direct product ${\Bbb
Z}_N\otimes{\Bbb Z}_N$ by the cyclic group $C_N$. \cite{poz1} This
group, denoted as above by ${\cal T}$, consists of elements
$(\omega^j,[k,l])$, where $\omega=\exp (2\pi{\rm i}/N)$ and
$j,k,l=0,1,\ldots,N-1$. The multiplication rule is given by the
following formula (all additions modulo $N$)
 $$
 (\omega^j,[k,l])(\omega^{j'},[k',l'])
 \;=\;(\omega^{j+j'+hkl'},[k+k',l+l'])\,.
 $$
 The parameter $h=0,1,\ldots,N-1$ labels inequivalent extensions and
corresponds to the magnetic field {\bf H} in Eq.\ (\ref{facm}). It is
evident that algebraic properties of this group depend on~$h$ or,
strictly speaking, on the greatest common divisor $\gcd(h,N)$ since
for $\gcd(h,N)=\gcd(h',N)$ groups labeled by $h$ and $h'$ are
isomorphic. In the further considerations we assume $h=1$ in order to
reduce a number of parameters and of different cases. It is worthwhile 
to mention that for $\gcd(h,N)>1$ the extension of ${\Bbb
Z}_N\otimes{\Bbb Z}_N$ by $C_{N/\gcd(h,N)}$ with the multiplication
rule parameterized by $h/\gcd(h,N)$ should be taken into account.

It follows from Eq.\ (\ref{irrep}) that irreps of ${\cal T}$ are labeled by
$\xi=0,1,\ldots,N-1$ corresponding to the irreps of $C_N$, i.e.\ we have
$\Xi(\omega^j)=\omega^{\xi j}$.  For each $\xi$ we have to find all
(inequivalent) projective representations $\Lambda^\xi$ of ${\Bbb
Z}_n\otimes{\Bbb Z}_N$. These representations satisfy the following
conditions: (i) a factor system of $\Lambda^\xi$ is given as [see Eq.\
(\ref{nuf})] 
 $$
 \nu^\xi([k,l],[k',l']) \;=\;\omega^{\xi kl'}\,;
 $$
 (ii) for a given factor system $\nu^\xi$ we have 
 $$
 \sum |\Lambda^\xi|^2=N^2
 $$
 (the sum is taken over all inequivalent projective irreps with the factor
system $\nu^\xi$). \cite{alt} It can be shown that for given $\xi$ there are
$\gcd(\xi,N)^2$ projective representations, each of dimension
$N/\gcd(\xi,N)$.  These representations are labeled by numbers
$\kappa,\lambda=0,1,\ldots,\gcd(\xi,N)-1$, corresponding to irreps of ${\Bbb
Z}_{\gcd(\xi,N)}\otimes{\Bbb Z}_{\gcd(\xi,N)}$. (Thus for given $\xi$ the
crystal lattice is `scaled' $N/\gcd(\xi,N)$ times.) To make a long story
short an actual form of matrix elements will not be discussed but only some
general properties will be presented. (In fact so used irreps are similar to
those considered by Brown \cite{brown} and Zak. \cite{zak1,zak2})

It follows from the previous considerations that the representations (vector
ones of ${\cal T}$ or projective ones of $T$) with $\xi>1$ describe the
movement of a particle with a charge $-\xi e$. Note that the periodic
boundary conditions imply that particles with charge $q$ and $q+N$ behave in
the same way. In particular it also applies to products of irreps: a product
of two representations labeled by $\xi_1$ and $\xi_2$, respectively,
decomposes into a sum of representations labeled by $\xi_1+\xi_2$ (modulo
$N$). Thus a system of two particles with charges $-\xi_1e$ and $-\xi_2e$
has total charge $-(\xi_1+\xi_2)e$. This relation follows from the form the
first factor in Eq.\ (\ref{irrep}): 
 $$
 (\Xi_1\otimes\Xi_2)(\omega^j)\;=\; \omega^{(\xi_1+\xi_2) j}\,.
 $$
 In particular, a square of the $N$-dimensional representation
$\Gamma^1$ (determined by the unique projective irrep $\Lambda$)
corresponds to a pair of electrons. A number of terms and the
multiplicity coefficients $f(\kappa,\lambda)$ in the decomposition 
 $$
  \Gamma^{\xi_1}_{\kappa_1,\lambda_1}\otimes
  \Gamma^{\xi_2}_{\kappa_2,\lambda_2}\;=\;\bigoplus_{\kappa,\lambda}
  f(\kappa,\lambda)\Gamma^{\xi_1+\xi_2}_{\kappa,\lambda}
 $$
 depend on the arithmetic relations between $\xi_1$, $\xi_2$ and $N$.
In the case $\Gamma^1\otimes\Gamma^1$  one obtains different results
for $N$ odd and even. In the first case the product decomposes into
$N$ copies of the (unique) representation $\Gamma^2$, since
$\gcd(2,N)=1$.  On the other hand, for $N=2M$ one has $\gcd(2,2M)=2$
and the considered product decomposes into a direct sum of
$M$-dimensional representations. There are 4 inequivalent such
representations and each of them appears $M$ times.

\section{Examples}
 If $N$ is a prime number then $\gcd(\xi,N)=1$ or $N$, and it is easy to
determine decomposition of each product:
 \begin{eqnarray*}
   \Gamma^0_{\kappa,\lambda}\otimes\Gamma^0_{\kappa',\lambda'}
     &=& \Gamma^0_{\kappa+\kappa',\lambda+\lambda'}\,;\\
   \Gamma^0_{\kappa,\lambda}\otimes\Gamma^\xi
     &=& \Gamma^\xi\,, \qquad \mbox{\rm for\ } 
     \xi=1,2,\ldots,N-1\,;\\
   \Gamma^\xi\otimes\Gamma^{N-\xi}
     &=& \bigoplus_{\kappa,\lambda=0}^{N-1}
     \Gamma^0_{\kappa,\lambda}\,;\\
   \Gamma^{\xi_1}\otimes\Gamma^{\xi_2}
     &=& N\Gamma^{\xi_1+\xi_2}\,,\qquad\mbox{\rm for\ }
     \xi_1+\xi_2\neq N\,.
  \end{eqnarray*}

The first nontrivial case corresponds to $N=4$. However, this case does not
show all the richness of possible products, since there is only one
nontrivial divisor $\xi=2$. The central extension of ${\Bbb Z}_4\otimes{\Bbb
Z}_4$ has 22 irreps:\\
 \hspace*{5mm} (i) 16 one-dimensional ones for $\xi=0$ labeled by
$\kappa,\lambda=0,1,2,3$; they are simply the ordinary vector
representations of ${\Bbb Z}_4\times{\Bbb Z}_4$;\\
 \hspace*{5mm} (ii) 2 four-dimensional ones for $\xi=1$ and
$\xi=3$;\\
 \hspace*{5mm} (iii) 4 two-dimensional ones for $\xi=2$ labeled by
$\kappa,\lambda=0,1$.

Two-electron states form a 16-dimensional space with the basis vectors
$|p_1p_2\rangle$, where $p_1,p_2=0,1,2,3$ label vectors of the
representation $\Gamma^1$. This space decomposes into 8 two-dimensional
representations $\Gamma^2_{\kappa,\lambda}$ with $f(\kappa,\lambda)=2$ for
all $\kappa,\lambda=0,1$. Hence, the irreducible basis can be denoted as
$|\kappa\lambda vq\rangle$, where $v=0,1$ is the repetition index and
$q=0,1$ labels vectors of $\Gamma^2_{\kappa,\lambda}$. The relatively simple
form of matrix elements allows determination of the Clebsch--Gordan
coefficients. In the presented case they lead to the following formulae:
 $$
\begin{array}{@{}l@{~~}l}
  |0000\rangle = (|00\rangle +|22\rangle)/\sqrt2, &
  |0001\rangle = (|11\rangle +|33\rangle)/\sqrt2, \\[-1pt]
  |0010\rangle =(|13\rangle +|31\rangle)/\sqrt2, &
  |0011\rangle = (|02\rangle +|20\rangle)/\sqrt2, \\[-1pt]
  |0100\rangle = {\rm i}(|00\rangle -|22\rangle)/\sqrt2, &
  |0101\rangle = (|11\rangle -|33\rangle)/\sqrt2, \\[-1pt]
  |0110\rangle = (|13\rangle -|31\rangle)/\sqrt2, &
  |0111\rangle = {\rm i}(|02\rangle -|20\rangle)/\sqrt2, \\[-1pt]
  |1000\rangle = (|01\rangle +|23\rangle)/\sqrt2, &
  |1001\rangle = (|12\rangle +|30\rangle)/\sqrt2, \\[-1pt]
  |1010\rangle = (|10\rangle +|32\rangle)/\sqrt2, &
  |1011\rangle = (|03\rangle +|21\rangle)/\sqrt2, \\[-1pt]
  |1100\rangle = {\rm i}(|01\rangle -|23\rangle)/\sqrt2, &
  |1101\rangle = (|12\rangle -|30\rangle)/\sqrt2, \\[-1pt]
  |1110\rangle = (|10\rangle -|32\rangle)/\sqrt2, &
  |1111\rangle = {\rm i}(|03\rangle -|21\rangle)/\sqrt2.
  \end{array}
$$

The numbers $p_1$ and $p_2$ can be interpreted as quasi-momenta since we
have $\widehat{T}({\bf a}_2)|p_1\rangle=|p_1-1\rangle$ [cf.\ Ref.~4
Eq.~(25)].  The translation along ${\bf a}_2$ is distinguished due to the
choice of the matrix form of the considered representations.  In general,
there is always one distinguished direction and the number $p$ labels the
corresponding quasi-momentum. \cite{zak3} Such interpretation of the indices
$p_1$ and $p_2$ allows the introduction of a Hamiltonian which commutes with
all operators $\Gamma^1\otimes\Gamma^1(u,{\bf R})$ (matrix elements are
given):
 $$
  {\cal H}_{p_1p_2,p'_1p'_2}\;=\; 
   \delta_{p_1+p_2,p'_1+p'_2} a_{p_1+p_2,p_1-p'_1}\;,
 $$
 where
 $$
 a_{0,0} \;=\; a_{2,0}\;=\;a_0,\quad a_{1,0}\;=\;a_{3,0}\;=\;a_1,
 \quad a_{p,1} \;=\; a_{p,3}.
 $$
 All these relations follow from the symmetry requirements. The terms
$a_{p,0}$ correspond to the total quasi-momentum $p$ and describe the
kinetic energy ($a_{p,0}>0$); the condition $a_{0,0}=a_{2,0}$ is connected
with `re-scaling' of the lattice since the representations $\Gamma^2$ are
two-dimensional. The terms $a_{p,q}$ for $q\neq0$ correspond to the
interchange of a quasi-particle with the quasi-momentum $q$ or, in the other
words, to the interaction of electrons. In the simplest approximation one
can assume that $a_{p,q}$ for $q\neq0$ does not depend on $p$ (so it will be
hereafter denoted as $b_q$; recall that $b_1=b_3$) and is negative. It is
also natural to assume that $a_0<a_1$ and that the probability of
interaction with $q=2$ is smaller than this one for $q=1$ (to begin with one
can assume $b_2=0$).

 In such an approximation one finds that levels corresponding to
$\Gamma^2_{01}$ and $\Gamma^2_{11}$ are four-fold degenerated with energies
$a_0-b_2$ and $a_1-b_2$, respectively. The representation $\Gamma^2_{10}$
leads to two 2-fold degenerated levels with energies $a_1+b_2\pm2b_1$.
Similarly, one obtains that two representations $\Gamma^2_{00}$ describe
levels with energies $a_0+b_2\pm2b_1$, respectively. In two later cases the
following linear combinations of vectors take the form:
 $$
 {1\over{\sqrt2}}(|\kappa000\rangle \pm |\kappa010\rangle)\,,
 \qquad\mbox{\rm for}\qquad \kappa\:=\:0,1\,.
 $$
 The ground-state energy is $E=a_0+b_2+2b_1$ and the corresponding
eigenvector is 
 $$
 {1\over2}(|00\rangle +|22\rangle+|13\rangle +|31\rangle)\,, 
 $$
 i.e.\ it is the sum of states $|p,-\!p\rangle$. Such a result resembles the
BCS state but it is not antisymmetric.  However, the performed
investigations are semi-classical and electrons have been considered as
spinless particles.

\section{Final remarks}
 The algebraic analysis of the magnetic translation groups (or,
equivalently, of the projective irreducible representations of the
translation group) gives us deeper insight into their structure. This
relates to many physical problems: movement of charged particles in a
magnetic (or an electromagnetic) field and a periodic potential, high
$T_c$-superconductors, the Hall effects (especially the fractional quantum
Hall effect), anyons, finite phase spaces etc.  The above presented
considerations indicate the importance of the product of representations.
The discussed examples are very simple and the physical interpretation is a
bit na\"\i ve, but they have shown the main (mathematical) properties of the
proposed picture.

Let $\varphi=hc/e$ be a fluxon and ${\bf H}={\bf h}\varphi$.
Replacing the electron charge $e$ by a charge $Q=-\xi e$ the factor
system (\ref{facm}) determined by Brown can be written as
 $$
  m({\bf R},{\bf R}')\;=\;\exp\Bigl[2\pi{\rm i}\,
  \xi\,\case{1}{2}({\bf R}\times{\bf R}')\cdot{\bf h} \Bigr].
 $$
 This formula shows that physical properties, which depend on this
factor, are periodic with respect to the magnetic field, lattice
vectors and the charge. The first case had been pointed out by Azbel
\cite{azbel} and noted also by Zak. \cite{zak2} The second is, in a
sense, the basis of introduction of magnetic cells
\cite{brown,zak1,zak,hald} ({\bf R} and ${\bf R}'$ are
linear combinations with integer coefficients of
basis vectors ${\bf a}_j$). This
work has shown that also the periodicity with respect to the charge
of a particle should be taken into account.

\end{document}